\RequirePackage{lineno}
\documentclass[twocolumn,showpacs,preprintnumbers,amsmath,amssymb,prl,superscriptaddress]{revtex4}
\usepackage{graphicx}
\usepackage{dcolumn}
\usepackage{bm}
\def\ra    {\rightarrow}

\def\Jpsi  {\ensuremath{J/\psi}}

\def\Bd    {\ensuremath{B^0}}
\def\Bu    {\ensuremath{B^+}}
\def\Bs    {\ensuremath{B_s^0}}
\def\bBs   {\ensuremath{\bar {B}_s^0}}
\def\BsJpsiPhi {\ensuremath{\Bs \ra J/\psi\,\phi}}

\def\BdJpsiKst {\ensuremath{\Bd \ra J/\psi\,K^{*0}}}
\def\BuJpsiK   {\ensuremath{\Bu \ra J/\psi\,K^+}}

\def\mevc  {\ifmmode {\rm MeV}/c \else MeV$/c$\fi}
\def\mevcc {\ifmmode {\rm MeV}/c^2 \else MeV$/c^2$\fi}
\def\gevc  {\ifmmode {\rm GeV}/c \else GeV$/c$\fi}
\def\gevcc {\ifmmode {\rm GeV}/c^2 \else GeV$/c^2$\fi}
\def\lxy   {\ifmmode L_{\rm xy} \else $L_{\rm xy}$\fi}
\def\dedx  {\ifmmode {\rm d}E/{\rm d}x \else d$E/$d$x$\fi}

\begin{document}

\title{\boldmath First Flavor-Tagged Determination of Bounds on Mixing-Induced $\textit{CP}$ Violation in $\BsJpsiPhi$ Decays}

\affiliation{Institute of Physics, Academia Sinica, Taipei, Taiwan 11529, Republic of China} 
\affiliation{Argonne National Laboratory, Argonne, Illinois 60439} 
\affiliation{Institut de Fisica d'Altes Energies, Universitat Autonoma de Barcelona, E-08193, Bellaterra (Barcelona), Spain} 
\affiliation{Baylor University, Waco, Texas  76798} 
\affiliation{Istituto Nazionale di Fisica Nucleare, University of Bologna, I-40127 Bologna, Italy} 
\affiliation{Brandeis University, Waltham, Massachusetts 02254} 
\affiliation{University of California, Davis, Davis, California  95616} 
\affiliation{University of California, Los Angeles, Los Angeles, California  90024} 
\affiliation{University of California, San Diego, La Jolla, California  92093} 
\affiliation{University of California, Santa Barbara, Santa Barbara, California 93106} 
\affiliation{Instituto de Fisica de Cantabria, CSIC-University of Cantabria, 39005 Santander, Spain} 
\affiliation{Carnegie Mellon University, Pittsburgh, PA  15213} 
\affiliation{Enrico Fermi Institute, University of Chicago, Chicago, Illinois 60637} 
\affiliation{Comenius University, 842 48 Bratislava, Slovakia; Institute of Experimental Physics, 040 01 Kosice, Slovakia} 
\affiliation{Joint Institute for Nuclear Research, RU-141980 Dubna, Russia} 
\affiliation{Duke University, Durham, North Carolina  27708} 
\affiliation{Fermi National Accelerator Laboratory, Batavia, Illinois 60510} 
\affiliation{University of Florida, Gainesville, Florida  32611} 
\affiliation{Laboratori Nazionali di Frascati, Istituto Nazionale di Fisica Nucleare, I-00044 Frascati, Italy} 
\affiliation{University of Geneva, CH-1211 Geneva 4, Switzerland} 
\affiliation{Glasgow University, Glasgow G12 8QQ, United Kingdom} 
\affiliation{Harvard University, Cambridge, Massachusetts 02138} 
\affiliation{Division of High Energy Physics, Department of Physics, University of Helsinki and Helsinki Institute of Physics, FIN-00014, Helsinki, Finland} 
\affiliation{University of Illinois, Urbana, Illinois 61801} 
\affiliation{The Johns Hopkins University, Baltimore, Maryland 21218} 
\affiliation{Institut f\"{u}r Experimentelle Kernphysik, Universit\"{a}t Karlsruhe, 76128 Karlsruhe, Germany} 
\affiliation{Center for High Energy Physics: Kyungpook National University, Daegu 702-701, Korea; Seoul National University, Seoul 151-742, Korea; Sungkyunkwan University, Suwon 440-746, Korea; Korea Institute of Science and Technology Information, Daejeon, 305-806, Korea; Chonnam National University, Gwangju, 500-757, Korea} 
\affiliation{Ernest Orlando Lawrence Berkeley National Laboratory, Berkeley, California 94720} 
\affiliation{University of Liverpool, Liverpool L69 7ZE, United Kingdom} 
\affiliation{University College London, London WC1E 6BT, United Kingdom} 
\affiliation{Centro de Investigaciones Energeticas Medioambientales y Tecnologicas, E-28040 Madrid, Spain} 
\affiliation{Massachusetts Institute of Technology, Cambridge, Massachusetts  02139} 
\affiliation{Institute of Particle Physics: McGill University, Montr\'{e}al, Canada H3A~2T8; and University of Toronto, Toronto, Canada M5S~1A7} 
\affiliation{University of Michigan, Ann Arbor, Michigan 48109} 
\affiliation{Michigan State University, East Lansing, Michigan  48824} 
\affiliation{University of New Mexico, Albuquerque, New Mexico 87131} 
\affiliation{Northwestern University, Evanston, Illinois  60208} 
\affiliation{The Ohio State University, Columbus, Ohio  43210} 
\affiliation{Okayama University, Okayama 700-8530, Japan} 
\affiliation{Osaka City University, Osaka 588, Japan} 
\affiliation{University of Oxford, Oxford OX1 3RH, United Kingdom} 
\affiliation{University of Padova, Istituto Nazionale di Fisica Nucleare, Sezione di Padova-Trento, I-35131 Padova, Italy} 
\affiliation{LPNHE, Universite Pierre et Marie Curie/IN2P3-CNRS, UMR7585, Paris, F-75252 France} 
\affiliation{University of Pennsylvania, Philadelphia, Pennsylvania 19104} 
\affiliation{Istituto Nazionale di Fisica Nucleare Pisa, Universities of Pisa, Siena and Scuola Normale Superiore, I-56127 Pisa, Italy} 
\affiliation{University of Pittsburgh, Pittsburgh, Pennsylvania 15260} 
\affiliation{Purdue University, West Lafayette, Indiana 47907} 
\affiliation{University of Rochester, Rochester, New York 14627} 
\affiliation{The Rockefeller University, New York, New York 10021} 
\affiliation{Istituto Nazionale di Fisica Nucleare, Sezione di Roma 1, University of Rome ``La Sapienza," I-00185 Roma, Italy} 
\affiliation{Rutgers University, Piscataway, New Jersey 08855} 
\affiliation{Texas A\&M University, College Station, Texas 77843} 
\affiliation{Istituto Nazionale di Fisica Nucleare, University of Trieste/\ Udine, Italy} 
\affiliation{University of Tsukuba, Tsukuba, Ibaraki 305, Japan} 
\affiliation{Tufts University, Medford, Massachusetts 02155} 
\affiliation{Waseda University, Tokyo 169, Japan} 
\affiliation{Wayne State University, Detroit, Michigan  48201} 
\affiliation{University of Wisconsin, Madison, Wisconsin 53706} 
\affiliation{Yale University, New Haven, Connecticut 06520} 
\author{T.~Aaltonen}
\affiliation{Division of High Energy Physics, Department of Physics, University of Helsinki and Helsinki Institute of Physics, FIN-00014, Helsinki, Finland}
\author{J.~Adelman}
\affiliation{Enrico Fermi Institute, University of Chicago, Chicago, Illinois 60637}
\author{T.~Akimoto}
\affiliation{University of Tsukuba, Tsukuba, Ibaraki 305, Japan}
\author{M.G.~Albrow}
\affiliation{Fermi National Accelerator Laboratory, Batavia, Illinois 60510}
\author{B.~\'{A}lvarez~Gonz\'{a}lez}
\affiliation{Instituto de Fisica de Cantabria, CSIC-University of Cantabria, 39005 Santander, Spain}
\author{S.~Amerio}
\affiliation{University of Padova, Istituto Nazionale di Fisica Nucleare, Sezione di Padova-Trento, I-35131 Padova, Italy}
\author{D.~Amidei}
\affiliation{University of Michigan, Ann Arbor, Michigan 48109}
\author{A.~Anastassov}
\affiliation{Rutgers University, Piscataway, New Jersey 08855}
\author{A.~Annovi}
\affiliation{Laboratori Nazionali di Frascati, Istituto Nazionale di Fisica Nucleare, I-00044 Frascati, Italy}
\author{J.~Antos}
\affiliation{Comenius University, 842 48 Bratislava, Slovakia; Institute of Experimental Physics, 040 01 Kosice, Slovakia}
% ...
\collaboration{CDF Collaboration\footnote{With visitors from 
$^a$University of Athens, 15784 Athens, Greece, 
$^b$Universidad Aut\'onoma of Madrid, E-28049 Madrid, Spain,
$^c$Chinese Academy of Sciences, Beijing 100864, China, 
$^d$University of Bristol, Bristol BS8 1TL, United Kingdom, 
$^e$University Libre de Bruxelles, B-1050 Brussels, Belgium, 
$^f$University of California Irvine, Irvine, CA  92697, 
$^g$University of California Santa Cruz, Santa Cruz, CA  95064, 
$^h$Cornell University, Ithaca, NY  14853, 
$^i$University of Cyprus, Nicosia CY-1678, Cyprus, 
$^j$University College Dublin, Dublin 4, Ireland, 
$^k$University of Edinburgh, Edinburgh EH9 3JZ, United Kingdom, 
$^l$University of Heidelberg, D-69120 Heidelberg, Germany, 
$^m$Universidad Iberoamericana, Mexico D.F., Mexico, 
$^n$University of Manchester, Manchester M13 9PL, England, 
$^o$Nagasaki Institute of Applied Science, Nagasaki, Japan, 
$^p$University de Oviedo, E-33007 Oviedo, Spain, 
$^q$Queen Mary, University of London, London, E1 4NS, England, 
$^r$Texas Tech University, Lubbock, TX  79409, 
$^s$IFIC(CSIC-Universitat de Valencia), 46071 Valencia, Spain, 
}}
\noaffiliation

\date{\today}

\begin{abstract}
This Letter describes the first determination of bounds on the
$\textit{CP}$-violation parameter $2\beta_s$ using $\Bs$ decays in
which the flavor of the bottom meson at production is identified.  The
result is based on approximately $2,000$ $\BsJpsiPhi$ decays
reconstructed in a 1.35~fb$^{-1}$ data sample collected with the
CDF\,II detector using $p\bar p$~collisions produced at the Fermilab
Tevatron.  We report confidence regions in the two-dimensional space
of $2\beta_s$ and the decay-width difference $\Delta\Gamma$.  Assuming
the standard model predictions of $2\beta_s$ and $\Delta\Gamma$, the
probability of a deviation as large as the level of the observed data
is 15\%, corresponding to 1.5 Gaussian standard deviations.

Dedicated to the memory of our dear friend and colleague, Michael
P. Schmidt. 
\end{abstract}

                             
\maketitle


The accurate determination of charge-conjugation-parity
($\textit{CP}$) violation in meson systems has been one of the goals
of particle physics since the effect was first discovered in neutral
kaon decays in 1964~\cite{Ref:Cronin}.  Standard model
$\textit{CP}$-violating effects are described through the
Cabibbo-Kobayashi-Maskawa (CKM) mechanism~\cite{Ref:Cabibbo}, which
has proved to be extremely successful in describing the phenomenology
of $\textit{CP}$ violation in $\Bd$ and $\Bu$ decays in the past
decade~\cite{Ref:PDG}.  However, comparable experimental knowledge of
$\Bs$ decays has been lacking.

In the $\Bs$ system, the mass eigenstates $B^0_{sL}$ and $B^0_{sH}$
are admixtures of the flavor eigenstates $\Bs$ and $\bBs$.  This
causes oscillations between the $\Bs$ and $\bBs$ states with a
frequency proportional to the mass difference of the mass eigenstates,
$\Delta m_s\equiv m_H - m_L$. In the standard model this effect is
explained in terms of second-order weak processes involving virtual
massive particles that provide a transition amplitude between the
$\Bs$ and $\bBs$ states. The magnitude of this mixing amplitude is
proportional to the oscillation frequency, while its phase,
responsible for $\textit{CP}$~violation in $\BsJpsiPhi$ decays, is
$-2\beta_s^{\textit{SM}} = -2\arg\left(-{V_{ts}V_{tb}^*\over
V_{cs}V_{cb}^*}\right)$~\cite{Ref:Nierste}, where $V_{ij}$ are the
elements of the CKM quark mixing matrix.  The presence of physics
beyond the standard model could contribute additional processes and
modify the magnitude or the phase of the mixing amplitude.  The recent
precise determination of the oscillation frequency~\cite{Ref:BsMixing}
indicates that contributions of new physics to the magnitude, if any,
are extremely small~\cite{Ref:Ligeti}.  Global fits of experimental
data tightly constrain the $\textit{CP}$~phase to small values in the
context of the standard model, $2\beta_s^{\textit{SM}}\approx
0.04$~\cite{Ref:HFAG}.  However, new physics may contribute
significantly larger values~\cite{Ref:Ligeti,Ref:Nierste2}.  The
observed $\textit{CP}$~phase can be expressed as $2\beta_s =
2\beta_s^{\textit{SM}} - \phi_s^{\textit{NP}}$, where
$\phi_s^{\textit{NP}}$ is due to the additional processes.  The
decay-width difference between the mass eigenstates,
$\Delta\Gamma\equiv\Gamma_L-\Gamma_H$, is also sensitive to the same
new physics phase.  If $\phi_s^{\textit{NP}}\gg
2\beta_s^{\textit{SM}}$, we expect $\Delta\Gamma =
2|\Gamma_{12}|\cos(2\beta_s)$~\cite{Ref:Nierste2}, where
$|\Gamma_{12}|$ is the off-diagonal element of the $\Bs$-$\bBs$ decay
matrix from the Schroedinger equation describing the time evolution of
$\Bs$ mesons~\cite{Ref:Kuhr}.  Recent studies of $\BsJpsiPhi$ decays
without identification of the initial flavor of the $\Bs$
meson~\cite{Ref:D0,Ref:Kuhr} have provided information on
$\Delta\Gamma$ and have some limited sensitivity to the
$\textit{CP}$~phase.

In this Letter we present the first study of the $\BsJpsiPhi$
decay~\cite{Ref:CC} in which the initial state of the $\Bs$~meson
({\it i.e.} whether it is produced as $\Bs$~or its anti-particle
$\bBs$) is identified in a process known as ``flavor tagging''.  Such
information is necessary to separate the time evolution of mesons
produced as $\Bs$ or $\bBs$.  By relating this time development with
the $\textit{CP}$~eigenvalue of the final states, which is accessible
through the angular distributions of the $\Jpsi$ and $\phi$ mesons, we
obtain direct sensitivity to the $\textit{CP}$-violating phase.  This
phase enters the time-development with terms proportional to both
$|\cos(2\beta_s)|$ and $\sin(2\beta_s)$.  Analyses of $\BsJpsiPhi$
decays that do not use flavor tagging are primarily sensitive to
$|\cos(2\beta_s)|$ and $|\sin(2\beta_s)|$, leading to a four-fold
ambiguity in the determination of $2\beta_s$~\cite{Ref:D0,Ref:Kuhr}.

This measurement uses 1.35~fb$^{-1}$ of data collected by the CDF
experiment at the Fermilab Tevatron between February 2002 and
September 2006.  The CDF\,II detector is described in detail in
Ref.~\cite{Ref:Bishai}.  Detector sub-systems relevant for this
analysis are described briefly here.  The tracking system is composed
of silicon micro-strip detectors surrounded by a multi-wire drift
chamber.  The drift chamber provides tracking information and charged
particle identification through the measurement of specific ionization
energy loss ($\dedx$).  A time-of-flight (TOF) detector provides
additional particle identification.  These detectors are immersed
within a 1.4~T axial magnetic field.  Electromagnetic and hadronic
calorimeters surround the solenoid.  At the outermost radial extent of
the detector, muons are detected in planes of multi-wire drift
chambers and scintillators.  The data used were collected with a
di-muon trigger which preferentially selects events containing
$\Jpsi\ra\mu^+\mu^-$ decays~\cite{Ref:Bishai}.

\begin{figure}[htbp]
\centerline{
  \makebox{
     \includegraphics[width=0.30\textwidth]{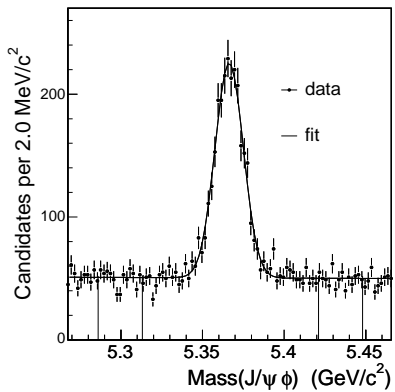}
   }
} 
\caption{\label{fig:mass}
Invariant $\mu^+\mu^-K^+K^-$ mass distribution with the fit projection
overlaid.  
The vertical lines indicate the mass sideband regions.}
\end{figure}

We reconstruct the $\BsJpsiPhi$ decay from the decays
$\Jpsi\ra\mu^+\mu^-$ and $\phi\ra K^+K^-$ and require these final
state particles to originate from a common point.  We use an
artificial neural network (ANN)~\cite{Ref:NN} to separate $\BsJpsiPhi$
signal from background.  In the ANN training, we consider the
following variables: particle identification of kaons using the TOF
and $\dedx$, the component of momenta of the $\Bs$ and $\phi$
candidates transverse to the proton beam direction, the invariant mass
of the $\phi$ candidate, and the quality of a kinematic fit to the
trajectories of the final state particles.  We have trained the ANN
with signal events from simulated data that are passed through the
standard {\sc geant}-based~\cite{Ref:Geant} simulation of the CDF\,II
detector~\cite{Ref:cdfsim} and are reconstructed as in real data.  We
use $\BsJpsiPhi$ mass sideband candidates, defined as those having
$m(\Jpsi\phi)\in[5.1820,5.2142]\ \cup\ [5.3430,5.3752]~\gevcc$, as the
background sample in the ANN training.  Applying the selection on the
output variable of the ANN, we observe $2,019 \pm 73$ $\BsJpsiPhi$
signal events with a signal to background ratio of approximately one.
The invariant $\Jpsi\phi$ mass distribution is shown in
Fig.~\ref{fig:mass}.  An event-specific primary interaction point is
used in the calculation of the proper decay time, $t = m(\Bs)
L_{xy}(\Bs)/p_T(\Bs)$, where $L_{xy}(\Bs)$ is the distance from the
primary vertex to the $\BsJpsiPhi$ decay vertex projected onto the
momentum of the $\Bs$ in the plane transverse to the proton beam
direction, $m(\Bs)$ is the mass of the $\Bs$~meson~\cite{Ref:PDG}, and
$p_T(\Bs)$ is its measured transverse momentum.

The orbital angular momenta of the vector (spin 1) mesons, $\Jpsi$ and
$\phi$, produced in the decay of the pseudoscalar (spin 0)
$\Bs$~meson, are used to distinguish the $\textit{CP}$-even S- and
D-wave final states from the $\textit{CP}$-odd P-wave final state.  We
measure the decay angles $\theta_T$, $\phi_T$, and $\psi_T$, defined
in Ref.~\cite{Ref:Kuhr}, in the transversity basis~\cite{Ref:Dighe}.
The transverse linear polarization amplitudes at $t=0$,
$A_{\parallel}$ and $A_{\perp}$, correspond to $\textit{CP}$~even and
$\textit{CP}$~odd final states, respectively.  The longitudinal
polarization amplitude $A_0$ corresponds to a $\textit{CP}$~even final
state.  The polarization amplitudes are required to satisfy the
condition $|A_0|^2 + |A_\parallel|^2 + |A_\perp|^2 = 1$.

In order to separate the time development of the $\Bs$~meson from that
of the $\bBs$~meson, we identify the flavor of the $\Bs$ or
$\bBs$~meson at the time of production by means of flavor tagging.
Two independent types of flavor tags are used, each exploiting
specific features of the production of $b$~quarks at the Tevatron,
where they are mostly produced as $b\bar{b}$~pairs.  The first type of
flavor tag infers the production flavor of the $\Bs$ or $\bBs$~meson
from the decay products of the $b$~hadron produced by the other
$b$~quark in the event.  This is known as an opposite-side flavor tag
(OST).  The OST decisions are based on the charge of muons or
electrons from semileptonic $B$~decays~\cite{Ref:Giurgiu,Ref:Tiwari}
or the net charge of the opposite-side jet~\cite{Ref:Lecci}.  If
multiple tags are available for an event, the decision from the
highest dilution flavor tag is chosen~\cite{Ref:BsMixing0}.  The tag
dilution ${\cal D}$, defined by the probability to correctly tag a
candidate $P_{tag}\equiv (1+{\cal D})/2$, is estimated for each event.
The calibration of the OST dilution is determined from $\BuJpsiK$ and
$\BdJpsiKst$ decays.  The second type of flavor tag identifies the
flavor of the reconstructed $\Bs$ or $\bBs$~meson at production by
correlating it with the charge of an associated kaon arising from
fragmentation processes~\cite{Ref:SSKT}, referred to as a same-side
kaon tag (SSKT).  The SSKT algorithm and its dilution calibration on
simulated data are described in Ref.~\cite{Ref:Belloni}.  The average
dilution is $(11\pm 2)\%$ for the OST and $(27\pm 4)\%$ for the SSKT,
where the uncertainties contain both statistical and systematic
effects.  The measured efficiencies for a candidate to be tagged are
$(96\pm 1)\%$ for the OST and $(50 \pm 1)\%$ for the SSKT.

An unbinned maximum likelihood fit is performed to extract the
parameters of interest, $2\beta_s$ and $\Delta\Gamma$, plus nuisance
parameters to the measurement, which include the signal fraction
$f_s$, the mean $\Bs$ width $\Gamma\equiv (\Gamma_L + \Gamma_H)/2$,
the mixing frequency $\Delta m_s$, the magnitudes of the polarization
amplitudes $|A_0|^2$, $|A_{\parallel}|^2$, and $|A_{\perp}|^2$, and
the strong phases $\delta_{\parallel}\equiv
\arg(A^*_{\parallel}A_0)$ and $\delta_{\perp}\equiv
\arg(A^*_{\perp}A_0)$. The fit uses information on the
reconstructed $\Bs$ candidate mass $m$ and its uncertainty $\sigma_m$,
the $\Bs$ candidate proper decay time $t$ and its uncertainty
$\sigma_{t}$, the transversity angles $\vec\rho = \{\cos\theta_T,
\phi_T, \cos\psi_T\}$, and tag information ${\cal D}$ and $\xi$, where
${\cal D}$ is the event-specific dilution and $\xi=\{-1,0,+1\}$ is the
tag decision, in which $+1$ corresponds to a candidate tagged as
$\Bs$, $-1$ to a $\bBs$, and $0$ to an untagged candidate.  The
single-event likelihood is described in terms of signal ($P_s$) and
background ($P_b$) probability distribution functions (PDFs) as
\begin{eqnarray}
          &&f_s P_s(m|\sigma_m) P_s(t,\vec\rho,\xi|{\cal D}, \sigma_{t}) 
               P_s(\sigma_{t}) P_s({\cal D}) \nonumber \\
          &&+(1-f_s) P_b(m) P_b(t| \sigma_{t}) P_b(\vec\rho) 
               P_b(\sigma_{t}) P_b({\cal D}).
\label{eqn:fullPDF}
\end{eqnarray}
The signal mass PDF $P_s(m|\sigma_m)$ is parameterized as a single
Gaussian with a standard deviation determined separately for each
candidate, while the background mass PDF, $P_b(m)$, is parameterized
as a first order polynomial.  The distributions of the decay time
uncertainty and the event-specific dilution are observed to be
different in signal and background, so we include their PDFs
explicitly in the likelihood.  The signal PDFs $P_s(\sigma_{t})$ and
$P_s({\cal D})$ are determined from sideband-subtracted data
distributions, while the background PDFs $P_b(\sigma_{t})$ and
$P_b({\cal D})$ are determined from the $\Jpsi\phi$ invariant mass
sidebands.  The PDFs of the decay time uncertainties,
$P_s(\sigma_{t})$ and $P_b(\sigma_{t})$, are described with a sum of
Gamma function distributions, while the dilution PDFs $P_s({\cal D})$
and $P_b({\cal D})$ are included as histograms that have been
extracted from data.

The time and angular dependence of the signal PDF
$P_s(t,\vec\rho,\xi,|{\cal D},\sigma_{t})$ for a single flavor tag
can be written in terms of two PDFs, $P$ for $\Bs$ and $\bar{P}$ for
$\bBs$, as
\begin{eqnarray}
  P_s(t,\vec\rho,\xi|{\cal D},\sigma_{t}) 
      &=& \frac{1+\xi{\cal D}}{2} P(t,\vec\rho| \sigma_{t}) \epsilon(\vec\rho)  \nonumber \\
      &&+\frac{1-\xi{\cal D}}{2}\bar{P}(t, \vec\rho| \sigma_{t}) \epsilon(\vec\rho),
\label{eqn:signalPDF}
\end{eqnarray}
which is trivially extended in the case of two independent flavor tags
(OST and SSKT).  The detector acceptance effects on the transversity
angle distributions, $\epsilon(\vec\rho)$, are modeled with
$\BsJpsiPhi$ simulated data.  Three-dimensional joint distributions of
the transversity angles are used to determine $\epsilon(\vec{\rho})$,
in order to correctly account for any dependencies among the angles.
The time and angular probabilities for $\Bs$ can be expressed as
\begin{eqnarray}
\frac{d^4 P(t,\vec\rho)}{dtd\vec\rho} 
&\propto& |A_0|^2 {\cal T}_+ f_1(\vec\rho) + |A_\parallel|^2 {\cal T}_+ f_2(\vec\rho) \nonumber \\
&+& |A_\perp|^2 {\cal T}_- f_3(\vec\rho) + |A_\parallel||A_\perp| {\cal U}_+ f_4(\vec\rho) \nonumber \\
&+& |A_0||A_\parallel| \cos(\delta_{\parallel}) {\cal T}_+ f_5(\vec\rho) \nonumber \\
&+& |A_0||A_\perp| {\cal V}_+ f_6(\vec\rho), \label{eqn:Bs} 
\end{eqnarray}
where the functions $f_1(\vec\rho)\dots f_6(\vec\rho)$ are defined in
Ref.~\cite{Ref:Kuhr}.  The probability $\bar{P}$ for $\bBs$ is
obtained by substituting ${\cal U_+}\ra{\cal U_-}$ and ${\cal
V_+}\ra{\cal V_-}$.  The time-dependent term ${\cal T}_{\pm}$ is
defined as
\begin{eqnarray*}
{\cal T}_{\pm}=e^{-\Gamma t}&\times&\left[\cosh(\Delta \Gamma t/2)
                 \mp \cos(2\beta_s)\sinh(\Delta \Gamma t/2)\right.\\
              & &\left.\mp\ \eta\sin(2\beta_s)\sin(\Delta m_st)\right], 
\end{eqnarray*}
where $\eta = +1$ for $P$ and $-1$ for $\bar{P}$.  The other
time-dependent terms are defined as
\begin{eqnarray*}
{\cal U}_{\pm} =\pm e^{-\Gamma t}&\times&\left[\sin(\delta_{\perp}-\delta_{\parallel})\cos(\Delta m_st) \right.\\
               &-&\cos(\delta_{\perp}-\delta_{\parallel})\cos(2\beta_s)\sin(\Delta m_st)                \\
               &\pm&\left.\cos(\delta_{\perp}-\delta_{\parallel})\sin(2\beta_s)\sinh(\Delta\Gamma t/2)
                  \right], \nonumber\\
{\cal V}_{\pm} =\pm e^{-\Gamma t}&\times&\left[\sin(\delta_{\perp})\cos(\Delta m_st) \right.\\
               &-&\cos(\delta_{\perp})\cos(2\beta_s)\sin(\Delta m_st)  \\
               &\pm&\left.\cos(\delta_{\perp})\sin(2\beta_s)\sinh(\Delta\Gamma t/2)\right].
\end{eqnarray*}
These relations assume that there is no direct $\textit{CP}$~violation
in the system.  The time-dependence is convolved with a Gaussian
proper time resolution function with standard deviation $\sigma_{t}$,
which is adjusted by an overall calibration factor determined from the
fit using promptly decaying background candidates.  The average of the
resolution function is 0.1~ps, with a root-mean-square deviation of
0.04~ps.

We model the lifetime PDF for the background, $P_b(t| \sigma_{t})$,
with a delta function at $t=0$, a single negative exponential, and two
positive exponentials, all of which are convolved with the Gaussian
resolution function.  The background angular PDFs are factorized,
$P_b(\vec\rho) = P_b(\cos\theta_T)P_b(\varphi_T)P_b(\cos\psi_T)$, and
are obtained using $\Bs$ mass sidebands events.

Possible asymmetries between the tagging rate and dilution of $\Bs$
and $\bBs$~mesons have been studied with control samples and found to
be statistically insignificant.  We allow important sources of
systematic uncertainty, such as the determination of overall
calibration factors associated with the proper decay time resolution
and the dilutions, to float in the fit.  The mixing frequency $\Delta
m_s = 17.77 \pm 0.12$~ps$^{-1}$ is constrained in the fit within the
experimental uncertainties~\cite{Ref:BsMixing}. Systematic
uncertainties coming from alignment, detector sculpting, background
angular distributions, decays from other $B$~mesons, the modeling of
signal and background are found to have a negligible effect on the
determination of both $\Delta\Gamma$ and $\beta_s$ relative to
statistical uncertainties.

An exact symmetry is present in the signal probability distribution,
as can be seen in Eq.~(\ref{eqn:Bs}), which is invariant under the
simultaneous transformation ($2\beta_s\to\pi-2\beta_s$,
$\Delta\Gamma\to -\Delta\Gamma$, $\delta_{\parallel}\to 2\pi -
\delta_{\parallel}$, and $\delta_{\perp}\to \pi -\delta_{\perp}$).
This causes the likelihood function to have two minima.  This symmetry
can be removed by restricting any of the above parameters within
appropriate ranges.  However, even after removal of the exact
symmetry, approximate symmetries remain, producing local minima.
Since the log-likelihood function is non-parabolic, we cannot
meaningfully quote point estimates. Instead we choose to construct a
confidence region in the $2\beta_s-\Delta\Gamma$ plane.

\begin{figure}[htbp]
\centerline{
  \makebox{
     \includegraphics[width=0.30\textwidth]{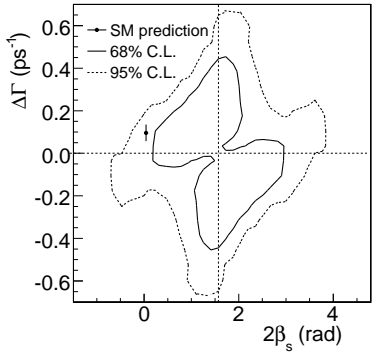}
   }
} 
\caption{\label{fig:contour}
Feldman-Cousins confidence region in the $2\beta_s-\Delta\Gamma$
plane, where the standard model favored point is shown with error
bars~\cite{Ref:Nierste2}.  The intersection of the horizontal and
vertical dotted lines indicates the reflection symmetry in the
$2\beta_s-\Delta\Gamma$ plane.}
\end{figure}

We use the Feldman-Cousins likelihood ratio ordering~\cite{Ref:FC1} to
determine the confidence level ($\textit{CL}$) for a $20\times40$ grid
evenly spaced in $2\beta_s\in[-\pi/2,3\pi/2]$ and
$\Delta\Gamma\in[-0.7,0.7]$.  The other parameters in the fit are
treated as nuisance parameters ({\it e.g.} $\Bs$ mean width,
transversity amplitudes, strong phases)~\cite{Ref:FC2}. The coverage
against deviations of the nuisance parameters from the measured values
is confirmed by randomly sampling the nuisance parameter space within
$\pm 5\sigma$ of the values determined from the fit to data.  The 68\%
and 95\% confidence regions obtained are shown in
Fig.~\ref{fig:contour}.  The solution centered in $0\leq 2\beta_s
\leq\pi/2$ and $\Delta\Gamma > 0$ corresponds to $\cos(\delta_{\perp})
< 0$ and $\cos(\delta_{\perp}-\delta_{\parallel}) > 0$, while the
opposite is true for the solution centered in $\pi/2 \leq \beta_s \leq
\pi$ and $\Delta\Gamma <0$.  Assuming the standard model predicted
values of $2\beta_s=0.04$ and $\Delta\Gamma = 0.096\mbox{
ps}^{-1}$~\cite{Ref:Nierste2}, the probability to observe a likelihood
ratio equal to or higher than what is observed in data is 15\%.
Additionally, we present a Feldman-Cousins confidence interval of
$2\beta_s$, where $\Delta\Gamma$ is treated as a nuisance parameter,
and find that $2\beta_s\in[0.32,2.82]$ at the 68\% confidence level.
The $\textit{CP}$~phase $2\beta_s$, $\Delta\Gamma$, $\Gamma$, and the
linear polarization amplitudes are consistent with those measured in
Ref.~\cite{Ref:Kuhr}.

We also exploit current experimental and theoretical information to
extract tighter bounds on the $\textit{CP}$-violating phase. By
applying the constraint $|\Gamma_{12}| = 0.048\pm
0.018$~\cite{Ref:Nierste2} in the relation $\Delta\Gamma =
2|\Gamma_{12}|\cos(2\beta_s)$, we obtain $2\beta_s\in [0.24,
1.36]\cup[1.78, 2.90]$ at the 68\% $\textit{CL}$.  If we additionally
constrain the strong phases $\delta_{\parallel}$ and $\delta_{\perp}$
to the results from $\BdJpsiKst$ decays~\cite{Ref:BaBar2} and the
$\Bs$ mean width to the world average $\Bd$~width~\cite{Ref:PDG}, we
find $2\beta_s\in[0.40, 1.20]$ at the 68\% $\textit{CL}$.

In summary we present confidence bounds on the $\textit{CP}$-violation
parameter $2\beta_s$ and the width difference $\Delta\Gamma$ from the
first measurement of $\BsJpsiPhi$ decays using flavor tagging.
Assuming the standard model predicted values of $2\beta_s=0.04$ and
$\Delta\Gamma = 0.096\mbox{ ps}^{-1}$, the probability of a deviation
as large as the level of the observed data is 15\%, which corresponds
to 1.5~Gaussian standard deviations.  Treating $\Delta\Gamma$ instead
as a nuisance parameter and fitting only for $2\beta_s$, we find that
$2\beta_s\in[0.32,2.82]$ at the 68\% confidence level.  The presented
experimental bounds restrict the knowledge of $2\beta_s$ to two of the
four solutions allowed in measurements that do not use flavor
tagging~\cite{Ref:D0, Ref:Kuhr} and improve the overall knowledge of
this parameter.


\begin{acknowledgments}
We would like to thank U.~Nierste for several useful suggestions.  We
thank the Fermilab staff and the technical staffs of the participating
institutions for their vital contributions. This work was supported by
the U.S. Department of Energy and National Science Foundation; the
Italian Istituto Nazionale di Fisica Nucleare; the Ministry of
Education, Culture, Sports, Science and Technology of Japan; the
Natural Sciences and Engineering Research Council of Canada; the
National Science Council of the Republic of China; the Swiss National
Science Foundation; the A.P. Sloan Foundation; the Bundesministerium
f\"ur Bildung und Forschung, Germany; the Korean Science and
Engineering Foundation and the Korean Research Foundation; the Science
and Technology Facilities Council and the Royal Society, UK; the
Institut National de Physique Nucleaire et Physique des
Particules/CNRS, France; the Russian Foundation for Basic Research;
the Comisi\'on Interministerial de Ciencia y Tecnolog\'{\i}a, Spain;
the European Community's Human Potential Programme; the Slovak R\&D
Agency; and the Academy of Finland.
\end{acknowledgments}


\bibliography{s2b_prl}    

\begin{thebibliography}{25}
\bibitem[{Ref({\natexlab{b}})}]{Ref:CC}
\bibinfo{note}{Charge-conjugate states are implied throughout the paper unless
  otherwise specified.}
\end{thebibliography}


\end{document}